\pdfoutput=1

\documentclass[11pt]{article}
\usepackage[utf8]{inputenc}
\usepackage{times}
\usepackage{subfig}
\usepackage{algorithm}
\usepackage[noend]{algorithmic}
\usepackage{nicefrac}
\usepackage{booktabs}
\usepackage[hang, flushmargin]{footmisc}
\usepackage[shortlabels]{enumitem}
\usepackage[margin=1in]{geometry}
\usepackage{comment}
\usepackage{booktabs,tabularx}
\usepackage{multirow}
\usepackage{svg}
\usepackage{textcomp}
\usepackage{listings}
\usepackage{authblk}

\usepackage[numbers]{natbib}
\usepackage{graphicx}
\usepackage{color}
\definecolor{darkgreen}{rgb}{0,0.5,0.0}
\definecolor{darkblue}{rgb}{0,0.0,0.5}
\definecolor{darkred}{rgb}{0.5,0.0,0.0}
\usepackage{hyperref}
\usepackage{wrapfig}
\usepackage{amssymb,amsmath,amsthm}
\usepackage[capitalise,noabbrev]{cleveref}
\usepackage{array}
\usepackage{url}
\usepackage[strict]{changepage}
\usepackage{makecell}
\usepackage{titlesec}
\usepackage{setspace}

\titleformat*{\subparagraph}{\itshape}

\newsavebox\actorsfigure

\title{Confidential Federated Computations}

\author{Hubert Eichner\footnote{H. Eichner and D. Ramage conceived, coordinated, and edited this work. Correspondence to \url{dramage@google.com}.}}
\author{Daniel Ramage\protect\footnotemark[1]}
\author{Kallista Bonawitz}
\author{Dzmitry Huba}
\author{Tiziano Santoro}
\author{Brett McLarnon}
\author{Timon Van Overveldt}
\author{Nova Fallen}
\author{Peter Kairouz}
\author{Albert Cheu}
\author{Katharine Daly}
\author{Adria Gascon}
\author{Marco Gruteser}
\author{Brendan McMahan}
\affil{Google Research}

\date{}

\begin{document}

\maketitle

\newcommand\blfootnote[1]{%
  \begingroup
  \renewcommand\thefootnote{}\footnote{#1}%
  \addtocounter{footnote}{-1}%
  \endgroup
}

\blfootnote{%
Change history:

- Version 2025-03: New \cref{sec:enhanced_external_verifiability} on enhanced external verifiability, and \cref{sec:confidential_federated_learning} on server-side federated learning. Various other clarifications.

- Version 2024-04: First release.

}

\begin{abstract}
Federated Learning and Analytics (FLA) have seen widespread adoption by technology platforms for processing sensitive on-device data. However, basic FLA systems have privacy limitations: they do not necessarily require anonymization mechanisms like differential privacy (DP), and provide limited protections against a potentially malicious service provider.
Adding DP to a basic FLA system currently requires either adding excessive noise to each device's updates, or assuming an honest service provider that correctly implements the mechanism and only uses the privatized outputs. Secure multiparty computation (SMPC) -based oblivious aggregations can limit the service provider's access to individual user updates and improve DP tradeoffs, but the tradeoffs are still suboptimal, and they suffer from scalability challenges and susceptibility to Sybil attacks.
This paper introduces a novel system architecture that leverages trusted execution environments (TEEs) and open-sourcing to both ensure confidentiality of server-side computations and provide externally verifiable privacy properties, bolstering the robustness and trustworthiness of private federated computations.

\end{abstract}



\setlength{\parskip}{0.5em}

\section{Introduction}
\label{sec:intro}
Since its introduction in 2017 \cite{mcmahan17fedavg, kairouz2019advances}, federated learning (FL) has seen adoption by technology platforms working with private on-device data (cross-device federated learning) or proprietary server-side data (cross-silo federated learning). FL's appeal has been driven by its straightforward privacy advantages: raw data stays in the control of participating entities, with only focused updates sent for immediate aggregation, visible to the service provider. Systems that realize federated learning \cite{bonawitz2019towards, huba2021papaya, paulik2021federated} run at scale today, reducing privacy risks in sensitive applications like mobile keyboards \cite{hard18gboard, yang18gboardquery, chen19oov, ramaswamy19emoji} and voice assistants \cite{apple19wwdc,hard2020training}.

However, basic federated learning offers an incomplete privacy story \cite{bonawitz2022federated}: updates sent to the service provider can reveal private data unless updates are aggregated obliviously, and aggregated updates can encode individual data unless trained with a differentially private (DP) learning algorithm \cite{dwork2006dp}. A dishonest service provider might log or inspect unaggregated messages, from which a great deal of information about an individual participant can be learned \cite{boenisch2023curious, suliman2023two}. This risk has been addressed with oblivious aggregation schemes that guarantee the service provider cannot inspect unaggregated messages, including secure multiparty computation (SMPC) from cohorts of honest devices \cite{bonawitz2017practical}, non-colluding SMPC-based secure aggregators \cite{talwar2023samplable}, or hardware trusted execution environments (TEEs) \cite{huba2021papaya}.  Recent work in training production FL models with device-level\footnote{For users with a single device, device-level DP corresponds directly to user-level DP.} DP has shown the potential of training high utility models with the strong guarantee that model parameters are statistically similar whether or not they contain any one device \cite{xu2024dpmfblog, xu2023federated}. And despite work showing how DP models can be trained atop oblivious aggregation protocols \cite{kairouz2021distributed, agarwal2021skellam}, substantial gaps remain between the best guarantees a model might offer with a trusted aggregator versus with SMPC-based secure aggregation. This is because state-of-the-art DP mechanisms either require sophisticated random device sampling schemes\footnote{For distributed DP implementations that use secure aggregation, leveraging random device sampling also requires hiding the secrecy of the sample. This can be done using private relays \cite{talwar2023samplable}.} or require statefulness \cite{choquette2024amplified, kairouz2021dpftrl}, making their distributed implementation challenging. This has led to utility gaps between central DP and distributed DP training\footnote{See figures 1 and 2 in \cite{kairouz2021dpftrl} where the performance of unamplified DP-SGD (achievable via distributed DP) is compared to the performance of stateful mechanisms like DP-FTRL.}. Further, robustness to Sybil attacks \cite{douceur2002sybil}, where the service provider sends specially constructed messages through the oblivious aggregator designed to maximize information about a targeted individual, continues to be a major challenge for SMPC-based secure aggregation schemes. 

In this paper, we introduce a new system architecture for federated learning and analytics \cite{fablog20}, designed to enable \emph{confidentiality of server-side computations and provide externally verifiable privacy properties}. In short, a long lived, TEE-hosted service holds the keys to uploaded encrypted data from client devices, and controls access to that data by TEE-hosted data processing pipelines via data access tracking and policy enforcement. TEE-hosted processing is subject to confidentiality, integrity, and external verifiability. The system enables a fleet of participating devices to keep control of their data, with individual uploads to the system tagged with a complete and closed set of computations in which the data may ever participate. Participating devices may further verify the privacy properties of the algorithms (e.g. a federated learning algorithm), as well as the properties of any unencrypted values released by the algorithm to the service provider (e.g., a specific DP guarantee). All uploaded messages and all intermediate results are encrypted with keys that are inaccessible to the service operator, so long as the TEEs' confidentiality and integrity properties hold. Our system design admits future opportunities to integrate SMPC as a defense against TEE failure or leakage, where applicable.

In order to prove to the device that uploaded data is released to the service provider only as part of a differentially private aggregate, our system is designed so that it can:
\begin{enumerate}
\item Prove to the device that the server is running binaries built from source code which exhibits the behaviors below, via remote attestation and open-sourcing - see \cref{sec:oss_repositories} for code pointers.
\item Prove to the device that uploaded data is processed only by a pre-declared series of operations, where intermediate results are encrypted, with the service provider only able to access explicitly released values. In \cref{sec:architecture} we describe the design of a \emph{ledger}, whose goal is to enforce this property.
\item Prove to verifiers that the pre-declared series of operations corresponds to an algorithm with an appropriate notion of privacy for the task at hand. We focus on the canonical case, where the algorithm computes a differentially private aggregate summary across many devices' data, with known DP properties, although other notions of privacy can be supported.
\item Prove to verifiers that the algorithm's privacy properties hold when it is run by the service provider across a collection of devices' input data. For DP algorithms, this includes correctness of the aggregation and accounting implementations, and the requirement that the service cannot re-run a segment of a DP algorithm multiple times with independent noise. Depending on the algorithm, it may also require proving that data is processed and combined under constraints such as shuffling or sampling.
\end{enumerate}

To the best of our knowledge, the system we have designed is the first to offer verifiable privacy in the sense outlined above. This enables new frontiers in private learning and analytics on federated data. For example, federated learning with verifiable DP guarantees that are robust to Sybil attacks can be implemented as a series of server-side processing steps that begin with per-client model updates. Similarly, models far larger than can be sent over the network, including LLM-sized models, can now be efficiently and privately trained on a TEE-hosted subset of federated data. We explore several more such applications, their privacy properties, and the design of the system in the sections that follow.

\section{Threat Model}
\label{sec:threat_model}
To precisely define the privacy and security guarantees of our proposed system, we begin by outlining anticipated threats and their potential impact. This involves identifying adversarial actors (both intentional and unintentional) and the technical capabilities they might employ. Recognizing that perfect security is unattainable, we then distinguish between threats our design can mitigate and those considered out of scope.

Our primary privacy objective is to offer proof to each end user (here: device owners) that the only way that their data can be used is in accordance with pre-approved privacy-preserving analyses.  Any other use by actors who might interact with our system constitutes a potential threat.  In general, our trust model considers adversaries who can span multiple of the roles below.

\subsection{Potential Adversaries and Capabilities}
\label{sec:potential_adversaries}

\paragraph{Data scientists} Any individual or group acting in the capacity of a data scientist is able to author queries for analyzing data or training machine learning models.  Inappropriately structured queries might not reflect the privacy properties we hope to enforce; for example, an inappropriate query might not offer enough (or any) differential privacy, which could conflict with the privacy promise made to the end user.  Such queries might be used to extract targeted information about individuals or groups. We require a system that offers verifiable proof that an analysis will \emph{only} be executed if it meets the required data privacy criteria.

Even if each analysis submitted by a data scientist is individually acceptable, the aggregate collection of analyses might not meet the desired privacy standards. Differential privacy, which mandates the addition of random noise, illustrates this challenge. If a data scientist can execute the same query repeatedly, each with fresh noise, they could potentially 'triangulate' the results (by averaging out the noise). This would allow them to infer the true, un-noised value with greater accuracy than a single differential privacy analysis suggests. Consequently, a system offering verifiable proof of the \emph{maximum number of times} a particular analysis can be executed is essential. 

\paragraph{Data access policy author}  Data access policies place codified limits on what computations (and how many) are allowed to run on user data, in particular the properties of the released outputs provided to the data scientist using the system. For example, the policy can specify specific DP parameters or aggregation requirements; when we refer to ``privacy guarantees'' or ``privacy properties'' in this work, we mean specifically those provided by an appropriate data access policy.
However, evaluating the acceptability of a data processing workload against a policy is only meaningful if the policy itself is well-defined, aligns with end-user expectations, and is resistant to unauthorized modification. Therefore, our system must ensure transparency regarding the access policy enforced upon data upload and provide verifiable proof that this policy cannot be covertly altered after data collection. That is, if the data access policy author is malicious or makes a mistake, the privacy guarantees of our system may not be adequate, but end users or auditors will be able reliably detect this shortcoming.

\paragraph{Other end users}  We require our system's privacy and security properties to be verifiable to each end user, regardless of the behavior of other end users.

\paragraph{Platform software author}  Server-side platform software authors have a great deal of flexibility over what code will interact with user data.  Intentional or unintentional deviations from expected behavior can have huge impacts on the resulting privacy, ranging from producing insufficient levels of differential privacy through directly exposing individual data.   In our system, we require that all code necessary to reason about the privacy properties of a given data processing workload is backed by verifiable proof of exactly what code has run.  Note that this may not be all the code in the system; for example, a portion of the system that schedules when to run an analysis may not need to be backed by verifiable proof, so long as the workload itself would meet the access policy whenever it does get run.

\paragraph{Platform software operator}  Any individual or group acting to operate the platform software also has significant capabilities to cause intentional and unintentional deviations from the data access policy, often in subtle ways.  For example, a platform operator might cause a single analysis to be executed multiple times, thereby circumventing expected DP protections as discussed above, under \emph{Data scientists}. 

The platform operator might also be able to control which users' data is supplied as input to which analyses, which can form the basis for several attacks.  For example, the differential privacy attack from the \emph{Data scientists} section supposes that different DP noise would be added each time an analysis is executed; might it be enough to guarantee that every time an analysis is executed, the \emph{same} DP noise is added?  In the presence of an adversary who can control which users' data is supplied to the analysis, the answer is \emph{no}: such an adversary could run the analysis once including a target user's data, a second time excluding that target user's data, and contrast the results (e.g. subtract one from the other) to recover just the target user's contribution.  We require a system that verifiably prevents this form of attack.

If a single end user's data can be copied many times and all the copies supplied to an analysis (as if they came from many users), then the analysis will be far more sensitive to the targeted users' data than expected; if DP is being used, the thus-inaccurate sensitivity analysis would lead to insufficient DP noise being added.  We require a system that prevents this kind of input replay attack.

A malicious platform operator might instead simulate many end users, using known data for each end user.  For example, if the analysis is intended to be a simple sum and the data access policy requires a certain number of users' data to be aggregated in that sum, then a malicious platform operator might include a single real users' data while simulating the rest of users as all holding the value zero---thereby producing a sum that is equal to the targeted user's data.  These \emph{Sybil} attacks \cite{douceur2002sybil} can make it particularly difficult to provide verifiable proof of privacy in the face of potentially malicious platform operators.  Properly implemented, central differential privacy protects each summand by clipping updates and adding noise, but this guarantee only protects against Sibyl attacks when an end user submitting their data to the system can verify that the analysis will be differentially private even if all other end users are actually fake or controlled by an adversary. Our system is designed to provide this property.

Many distributed systems include capabilities to persist or otherwise externalize portions of their operating state.  For example, a system might support some form of state snapshot that could enable recovering partial progress in the event that a process unexpectedly terminates, migrating an executing process from one machine to another, pausing and later resuming the process, and so on.  These capabilities and others can be extremely valuable, but can also lead to various kinds of \emph{rollback} attacks: where a portion of the platform system is forced to return to a previous state, either by changing the state of a live piece of the system, or by pausing and resuming from a previous state.  These capabilities can sometimes also lead to \emph{forking} attacks, where multiple copies of a piece of the system begin executing from some previously valid state, but wherein each copy behaves as if it were the only copy.  Forking and rollback attacks can lead to wide ranges of privacy-compromising behavior, often by enabling previously discussed attacks.  For example, if a system can produce an analysis, then be rolled back to a state from before the analysis was performed, it can be compelled to produce the analysis multiple times.  Alternatively, if the system can be forked before the analysis is performed, then each fork (behaving as if it were the only copy of the system) might perform the analysis separately.  As such, we require our system to provide verifiable proof against forking and rollback attacks, despite the careful consideration this will require in a robust and scalable distributed system.

\paragraph{Data center operator}  The platform software operator might choose to deploy the system on hardware directly controlled by a third party, for example Amazon AWS or Google Cloud; this is the party that directly supplies and operates the TEEs at the core of our system. Such data center operators have full access to the platform infrastructure in software, and could potentially execute many of the attacks above, including attempting to cause unexpected software to execute, dynamically controlling the network topology, starting / stopping / migrating workloads from one machine to another, and so on.  A data center operator typically runs the host operating system on which the platform workloads will run, controls the hypervisor on that system, and can observe various aspects of interaction between the analysis platform and the rest of the system.  The data center operator also has direct access to the physical system hardware, including the CPU and memory.

Our system assumes access to potentially insecure data center storage for (encrypted) intermediate state of the private computations. Similarly to the platform software operator, data center storage operators could attempt to read, modify, delete, inject, or rollback any data persisted into their storage systems.  We require our system to resist any such attack, e.g. by using various cryptographic measures to protect the confidentiality and integrity of any externalized data

Our goal is to provide defense against the potential attacks above under the assumption that the confidentiality and integrity properties of TEEs are maintained.  Specifically, we consider attacks based on physical access to the TEE hardware (e.g., power cycling, bus sniffing, etc.) to be out of scope, as the attacks require a very high degree of sophistication and there is no known mitigation for them. 
Side channel attacks (e.g., monitoring memory access patterns, timing of operations, message sizes, etc.) offer an interesting line of research that is out of scope in the current system, but will be revisited in future work---see the discussion in \cref{sec:side_channel_attacks}.
Finally, potential TEE failure due to vulnerabilities in the CPU or the hosted software \cite{schluter2024wesee} constitute known threats that a data center operator may be able to exploit. These also remain out of scope for the first version of our system presented here, which is thus not fully robust to a malicious data center operator in the presence of such vulnerabilities. We aim though for the design to be amenable to the use of secure multi-party computation and related techniques that could address such attacks going forward.

Due to these known failure modes of TEE technology, we consider the combination of TEE technology with Secure Multiparty Computations an important future direction (see \cref{sec:open_problems}).

\subsection{Approach: Distributed Trust, Trusted Execution Environments, and Hardware Manufacturers}
\label{sec:distributed_trust}

Defense against the attacks within the scope of our threat model become tractable when we are able to distribute trust between multiple parties wherein non-collusion between those parties is a believable assumption.  There exist many mechanisms for distributing trust; for example, in the cryptographic literature, we see SMPC as one such mechanism.

In this paper, we will be using hardware-based Trusted Execution Environments as the basis of our distributed trust.  In essence, we will distribute trust between the TEE hardware designer/manufacturer and remainder of the platform (composed of the platform software author and operator, and the data center storage and hardware operators). 
Specifically, we argue that that so long as \emph{either} the hardware manufacturers' guarantees hold or the remainder of the platform is uncompromised\footnote{E.g. because standard best practices for software systems security have been enforced.}, the privacy claims hold.  The privacy properties of our system will depend only on the code running inside the TEEs; any compromise of this code will either be publicly detectable (due to the external verifiability properties of the TEEs), or else will require both the TEE to fail to deliver at least one of its confidentiality / integrity / verifiability promises \textit{and} require an additional compromised operator to actively exploit that vulnerability.

\subsection{Related Work}
\label{sec:related_work}
Federated learning and analytics (FLA) have been the subject of extensive research. In this paper, we focus specifically on prior work relevant to large-scale FLA infrastructure and the integration of trusted computing techniques with differential privacy.

Privacy enforcement is at the heart of private data processing, and ideally the whole mechanism is both confidential and verifiable. Apple \cite{paulik2021federated} relies on encryption with ephemeral per round keys and in-memory aggregation with DP; however, neither client nor server side portions have been open-sourced. Papaya \cite{huba2021papaya} leverages TEEs to implement secure aggregation through random additive noise based on user supplied encrypted random seeds, but does not provide external verifiability. \citet{bonawitz2019towards} uses secure aggregation and has extensively open-sourced infrastructure \cite{fcpOnGithub}, but does not provide external verifiability beyond open-sourcing, and secure aggregation does not offer protections against Sybil attacks.

ElephantDP \cite{jin2024elephants}, unlike previous systems, provides a fully confidential and verifiable privacy budget mechanism and relies on the Narrator architecture \cite{narrator}, which provides state continuity via a distributed system  similar to ROTE \cite{rote} or Nimble \cite{nimble}, to enable fault tolerance and rollback protection. The ElephantDP system is closest to our work in that it enables running a series of DP queries over private data in a TEE, in a manner resistant to rollback or forking attacks. The main contributions of our system as compared to ElephantDP are the following. We consider a setting in which there is not a single data owner, but many client devices, which must all upload data in a secure yet scalable manner. We also consider a setting in which the compute required to run the DP queries is higher than that which can execute on a single TEE in a reasonable timeframe. Finally, we consider the impact of running queries containing some proprietary code on the privacy properties of the system.

\section{Architecture}
\label{sec:architecture}
In this section we first outline the design principles that motivate our architectural choices. We then proceed to describe the dataflow and various system components of our confidential federated computation platform.

\subsection{General Design Principles}
\label{sec:general_design}

\paragraph{Data anonymization} The results released by the system to the ``untrusted'' space are subject to data anonymization. We choose DP as the gold standard, but other application workload-appropriate anonymization methods can be supported.

\paragraph{Data minimization} Throughout the lifetime and placement of sensitive data, from data collection to aggregation, data minimization techniques are applied to limit as far as possible what data exists where, when, and who has access to it. This encompasses a wide range of techniques, from focused data collection, reducing the amount of collected data to the necessary minimum, to consistent encryption of non-DP-aggregated data.

\paragraph{Defense in depth} A secure data processing system should rely on multiple security mechanisms in case one or more of these mechanisms are breached. Throughout the stack we therefore employ various security best practices to minimize the risk of privacy leakage (some of these embody data minimization):
\begin{enumerate}
\item \textbf{Principle of least privilege}---limit the level of access, and the set of individuals or components with that access to the minimum necessary to operate the system. This is implemented through a variety of mechanisms such as \emph{Access Control Lists}. A notable application of this principle is minimizing the API surface between trusted and untrusted components to prevent attacks on trusted components, and leakage of private data from compromised trusted components.
\item Limit unilateral access to resources via \textbf{Multi-party authorization}, thus elevating the risk from individual malicious or compromised actors to multiple \emph{colluding} actors. This is further strengthened by \textbf{Separation of Duties}, the concept of assigning individuals to non-overlapping roles required to fulfill a task (e.g. ensure that data scientists and datacenter operators are not the same individual).
\item Increase \textbf{discoverability} of attacks through logging accesses and (manual or automated) auditing---a reactive method designed to act as a deterrent.
\item Employ \textbf{safe defaults} wherever possible, such as automatic data deletion after defined retention periods.
\end{enumerate}

\subsection{Trusted Computing Specific Design Principles}
\label{sec:trusted_computing}

\paragraph{Trusted vs. untrusted space} A well defined \textbf{decision boundary} for what functionality executes in trusted and in untrusted space is key to scalable and verifiable application of trusted computing in data processing systems. We follow the design philosophy of reducing TEE-hosted functionality to the logic necessary for providing privacy guarantees---notably, code paths that handle unanonymized user data. This approach:
\begin{enumerate}
\item Limits the attack surface of TEE-hosted software.
\item Allows for improved composition with existing, untrusted infrastructure for e.g. data storage, transport, and orchestration of computations; and further facilitates portability / reuse in other operating environments.
\item Reduces operating costs / improves performance due to TEE overhead (see \cref{sec:tee_overhead}).
\item Aids auditability by reducing the trusted computing base---notably the amount and scope of software components that need to be analyzed.
\end{enumerate}

\paragraph{External verifiability} One of our key motivations for using TEEs is to provide \textbf{verifiability} of claims about private data processing on the server side to outside parties. In particular, it should be easy for external auditors to verify or falsify our claims. For a list of properties we would like to prove, please see \cref{sec:intro}. Here we summarize prerequisites and technologies that we consider key to providing this verifiability:
\begin{enumerate}
\item Proving \emph{what} source code is being run in a TEE, requiring remote attestation to client devices and their users on what binaries are being run in a TEE, with the hardware (e.g. CPU) manufacturer being the root-of-trust. We present two approaches in \cref{sec:enabling_external_verifiability} and \cref{sec:enhanced_external_verifiability}.
\item Establishing what source code this binary refers to, via open-sourcing the code, publishing build instructions (see \cref{sec:oss_repositories}), build provenance, and reproducible builds (alternative---use of \emph{trusted build infrastructure}).
\item Provide high readability of the published code. While reducing Trusted Computing Base \cite{rushbyTCB} code size (which in this context means the code that is included in the binary running in the TEE) and complexity are good general guidelines, these steps may be at odds with flexibility and performance (e.g. consider executing high-performance ML workloads in a TEE); furthermore, the reuse of extensively studied infrastructure such as the Linux kernel may improve readability / verifiability over e.g. custom kernels with smaller TCB size but little to no existing auditing.
\end{enumerate}

At this time, we use TEEs to provide verifiability of the privacy sensitive code paths we execute on the \emph{server side}. On the device side, we face the challenge that TEEs are not widely available, and that the code needs to be integrated into existing apps that may be proprietary or built non-reproducibly by third parties; we therefore continue to provide device side transparency through open-sourcing the code (see \cref{sec:oss_repositories}).

\paragraph{Chain-of-trust} Most real world, large scale applications of TEEs for data processing will require distributing the work over multiple TEEs both to provide a) fault-tolerance for stateful components such as key management, and b) parallelization and pipelining for processing large amounts of data. Rather than having devices remotely attest each TEE involved in this distributed system, we propose the concept of a \emph{chain of trust}---devices establish trust with a specific binary, and that binary in turn will establish trust (through remote attestation, much like end user devices) with other binaries that will further handle the data, and so on. This is a useful abstraction mechanism to combine multiple binaries and TEEs into a distributed system while retaining full external verifiability of that system; our architecture presented in the next section will help illustrate this idea.

\subsection{Overview}
\label{sec:overview}
We begin our architectural description with a brief outline of the lifecycle of a trusted data processing pipeline, illustrated by describing the flow of data originating from end user devices through the system up to release of an anonymous aggregate, briefly touching on each of the involved components. We will then describe each component in greater detail in dedicated sections.

The first step is starting up a server-side trusted component we call the \emph{ledger}---a long lived service that literally ``holds the keys'' to uploaded encrypted data from client devices, and controls access to that data by other trusted components via data access tracking and policy enforcement.

Client devices store data locally, and retrieve three artifacts from the server---a query, a data access policy, and an attestation report from the ledger (including its public key). The device tests whether the data access policy---a description of how query results will be processed by server side pipelines---conforms to its settings (e.g. user preferences), including what computations can process the data, what the privacy budgets are etc. The device also verifies the attestation report; only after passing these checks, the data is encrypted with the ledger's public key and uploaded to server side storage.

Separately, a data processing pipeline such as a Map-Reduce style histogram computation is executed on the server. This pipeline orchestrates (spins up, chains) TEE-hosted data transformations to operate on the uploaded private data; these workloads request decryption keys from the ledger which may approve or deny the request subject to the access policy cryptographically bound to the data, and records the usage. Communication between these workloads is also encrypted. Finally, a data transformation workload may release aggregate results to untrusted space subject to differential privacy guarantees.

\begin{figure}
    \centering
    \includegraphics[width=0.85\textwidth]{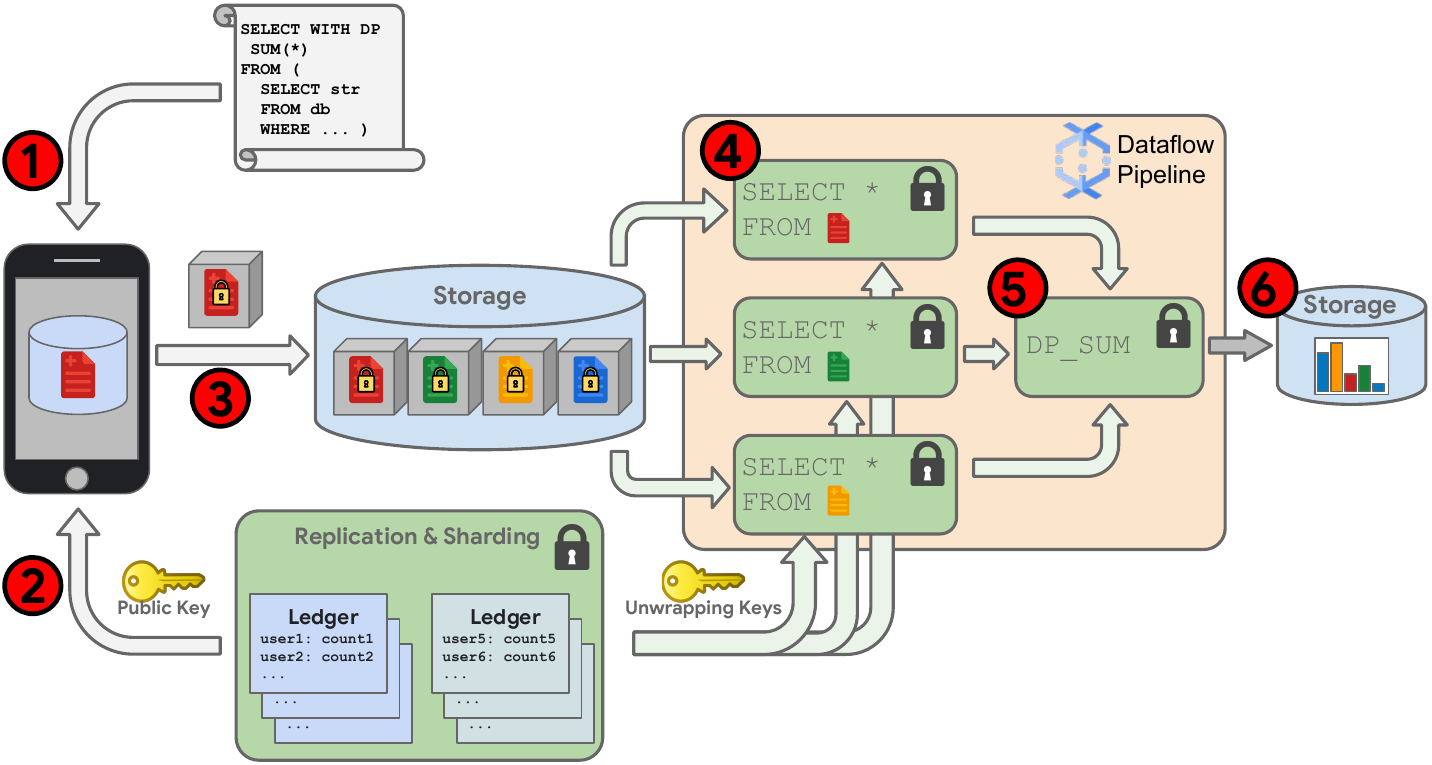}
    \caption{High level overview of client, TEE, and server data process in proposed architecture.}
    \label{fig:overview}
\end{figure}

\subsection{Trusted Infrastructure Primer}
\label{sec:trusted_infrastructure}

In order to be able to make externally verifiable claims about what happens to user data once it reaches the TEE, we need to be able to prove that the code running in the TEE was built from a specific open-source commit, without the need to ``just take our word for it.'' For this, we leverage Project Oak \cite{oakOnGithub}, a framework that provides hardware-based remote attestation bound to signing and encryption keys, and a layered architecture that allows verifying the entirety of the software stack running in the TEE.

At the bottom layer, a TEE usually only provides a single remote attestation primitive that binds together the measurement of the initial state of the TEE VM memory and some application-provided data. The measurement of the initial state of the VM corresponds to the digest of the compiled firmware binary (e.g. BIOS) with which the VM is booted. Oak provides a custom firmware (called ``stage0'') that allows booting either Linux or a custom kernel; Oak stage0 is intentionally very minimal (fully auditable by a single expert reviewer), written as much as possible in Rust, and reproducibly buildable from the Oak git repository. This allows external reviewers to confirm that the initial TEE measurement indeed corresponds to the specific code that they can read and audit on the Oak repository. Oak then uses the DICE protocol \cite{tcgDiceAttestationArchitecture} to recursively measure the rest of the boot chain, including the kernel, up to the application. At this time, we build on top of AMD SEV-SNP \cite{amdSEVSNP}, and plan to support Intel TDX \cite{intelTDX} in the future.

At the higher levels in the stack, Oak offers two alternatives:
\begin{itemize}
\item \textbf{Oak Restricted Kernel:} for high security use cases (with a minimal TCB, which does not include Linux), but relatively low performance and high migration overhead
\item \textbf{Oak Containers:} for high performance use cases, easy migration of existing workloads, access to hardware accelerators, at the cost of a larger TCB (including a Linux kernel and distribution)
\end{itemize}

Oak is implemented as a split architecture, in which only the subset of the logic that needs to be trusted runs in the TEE (and therefore contributes to the TCB) and has access to the decryption keys, while the rest of the logic that only deals with encrypted data and does not have access to the decryption keys runs outside of the TEE. The trusted and untrusted components communicate with each other over a communication channel in the form of a virtual network socket. This allows better separation of concerns, and only granting privileges to the minimal amount of logic that requires it.

The client (e.g. Android phone) obtains all this information from the server, and uses it to determine whether the server instance is trustworthy. There are three key pieces of information that it uses to this end (based on the RATS architecture \cite{rfc9334} terminology):
\begin{itemize}
\item \textbf{evidence:} claims signed by the TEE instance, binding together the state of the TEE, the identity of the code running in it, and application level data (usually public keys); produced by the TEE instance at runtime.
\item \textbf{endorsements:} additional data used to establish the trustworthiness of the evidence (e.g., intermediate certificates, signatures of binary hashes, transparency log inclusion proofs); produced or cached by the untrusted host running the TEE.
\item \textbf{reference values:} the policy that determines the acceptable range of values that the client trusts (e.g., TCB version number, individual binary hashes, signing keys with which binary hashes can be endorsed, etc.); usually hardcoded in the client code, or obtained over some trustworthy channel.
\end{itemize}

Oak provides a client library that allows a client to determine the trustworthiness of the TEE evidence according to a local set of reference values, taking into account any available endorsements. If the client is satisfied with the TEE evidence, it then extracts the public keys of the application, and uses them to verify signatures from and / or encrypt data to the application.

\subsection{Ledger}
\label{sec:ledger}
The ledger is the TEE-hosted application that enables the device, after verifying the identity of software in the TEE, to reason about the full set of downstream uses in which any uploaded data may participate. Conceptually, the ledger combines a \emph{key store} (holding keys needed to decrypt uploaded data and intermediate, unreleased state) with a \emph{stateful policy engine} (that enforces that keys are given out only to other TEE-hosted applications with approved access).

Specifically, the ledger is a key unwrapping service that enforces stateful use-based privacy policies. This service allows secure asynchronous communication between data producers and consumers using pre-existing channels, such as uploads with HTTP POST or processing with Apache Beam \cite{apacheBeam}. At a high level, a publisher---in our case, a user device or subsequently a TEE in a data processing pipeline---encrypts its data using authenticated encryption with associated data (AEAD) with an ephemeral key, wraps the ephemeral key with a ledger-generated public key using single-shot hybrid public key encryption (HPKE), then transmits the encrypted data and wrapped key using a channel of its choice. To decrypt the data, a consumer---in our case, a TEE in a data processing pipeline---provides the wrapped key and the identity of its TEE-hosted data transformation; if the consumer is authorized under the data's access policy, the ledger unwraps the key and returns it to the TEE-hosted data transformation over a secure channel. Unless that transformation is a terminal step in the data processing pipeline, it will then act as a data producer and re-encrypt the processed data before exposing it outside the TEE (see \cref{sec:data_processing}). The ledger maintains multiple HPKE keys so that keys can be rotated without turning up or down servers; these HPKE keys are signed by an application-level key in the Oak evidence to bind them to the TEE instance that generated them.

There are several important consequences of this system:
\begin{itemize}
\item Since the ledger only performs key unwrapping, its load does not scale with the size of the encrypted data.
\item Data producers do not need to verify every data transformation workload in the processing pipeline themselves. Instead, they can verify the ledger and the data access policy, entrusting verification of subsequent data access to the ledger. As a result, data producers and consumers do not need to run concurrently.
\item Data producers do not need to directly communicate with the ledger. Instead, they can obtain the ledger-generated public key, the ledger's evidence and endorsements, and the data access policy from an untrusted intermediary such as a Content Delivery Network. As a result, the ledger is not in the critical path for data production, relaxing its latency and availability requirements.
\item The wrapped key can only be unwrapped if the ledger still holds the corresponding HPKE private key. Cryptographic erasure of uploaded data can be achieved by intentionally deleting the HPKE private key. Additionally, when the HPKE private key is deleted, all associated data access records can safely be forgotten as well, bounding the amount of state that needs to be tracked. Wrapping the AEAD key for encrypted derived data with the same HPKE key as the source data ensures that the derived data cannot outlive the cryptographic erasure of the source data.
\item Separating data access policy enforcement from data processing allows existing processing systems to be augmented with external verifiability.
\end{itemize}

\subsubsection{Data Access Policies}
\label{sec:data_access_policies}
A \textit{data access policy} specifies a specific sequence of TEE-hosted transformations that can access uploaded data. Devices indelibly bind their encrypted uploads to a data access policy that fully determines the closed set of processing steps the data may partake in, as enforced by the ledger. The data access policy is uploaded as associated data (AD) used when wrapping the encrypted data's encryption key with HPKE. The access policy can be represented as a directed graph where each edge describes access conditions such as the specific TEE-hosted binary (and other components in the TEE including kernel and system image) as well as any privacy- or security-relevant aspects of the binary's configuration (e.g. $(\varepsilon, \delta)$ values for a DP aggregator). By combining the specific set of binaries that are able to access the data with knowledge of their (open) source code, it is possible to reason about the overall server-side usage of the data.

\begin{figure}
    \centering
    \includegraphics[width=0.35\textwidth]{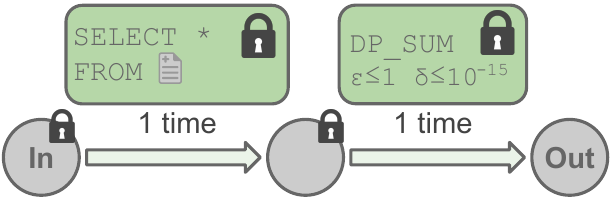}
    \caption{An example access policy for the pipeline in figure \ref{fig:overview} allowing the initial data to be processed first by a TEE that performs a SELECT operation, then by a TEE that implements a DP sum with $\varepsilon \leq 1$ and $\delta \leq 10^{-15}$. Each stage may only access the data once, and no other access is allowed. Data (gray) and transformations (green) with a ``lock'' icon are inaccessible to the system operator, assuming correct ledger and TEE functionality.}
    \label{fig:access_policy}
\end{figure}

Data access policies are stateful: they specify not only which consumers can access the data but also how many times (e.g. only once). As described in \cref{sec:threat_model}, limiting usage allows reasoning about the differential privacy properties of the system and is necessary to prevent denoising via triangulation. Tracking past data accesses significantly increases the amount of state maintained by the ledger, but by colocating that state with the corresponding HPKE private key, we ensure that the private key and access records have the same lifespan: attempting to wipe out access history by restarting the ledger would also wipe out the private key, rendering the associated client data unreadable. This state also needs to be resistant to forking and rollbacks, which is straightforward when the state is maintained in the memory of a single server and more involved in a multi-server or fault-tolerant environment (see \cref{sec:scalability_and_robustness}).

\subsubsection{Limiting Data Retention via Cryptographic Erasure}
\label{sec:ttls_via_cryptographic_erasure}
HPKE keys generated by the ledger have expiration times after which the ledger will forget about the private key and all associated access records---thus crypto-erasing all uploaded data protected by that key. However, as we will describe in \cref{sec:externally_verifiable_ttls}, accessing a secure time source from within a TEE is difficult, especially at high QPS. So instead of using a secure time source, the ledger uses an insecure but monotonically increasing time source that is exposed to clients in the \texttt{issued\_at} and \texttt{expiration} times of all generated HPKE keys. Client devices will not upload data if the ledger's time is too skewed: if the ledger's clock is running significantly behind the current time, its HPKE keys will already be expired, and if the ledger's clock is running significantly ahead of the current time, the \texttt{issued\_at} time will not yet be reached. See \cref{sec:externally_verifiable_ttls} for potential future improvements in this space.

To provide defense in depth, all encrypted data is also subject to storage-level time-to-live limits (TTLs), though these are not externally verifiable.

\subsection{Client Side Architecture}
\label{sec:client_side_architecture}
Clients (end user devices) participate in a confidential computation using a multi-step process that typically starts by collecting and storing data into an on-device database over a period of time; then, after checking in with the server, they perform a data summarization and result upload step. In cases where further data minimization is desirable or required, persisting data on devices may be skipped.

\subsubsection{Client Summarization Tasks}
\label{sec:client_summarization_tasks}
When a client connects to the server, the server provides it with a list of all currently available client summarization tasks, along with a set of optional \emph{eligibility criteria} describing the client requirements for each task. Summarization tasks describe how clients should preprocess and condense data before uploading results; in our current system, we support SQL queries that select a filtered projection of the on-device database, and machine learning workloads that compute model updates and metrics on that data. This preprocessing summarization offers flexibility and embodies the data minimization principle.

Before executing a summarization task, clients first determine whether they're eligible to run it by inspecting their local state using the aforementioned eligibility criteria. For example, \emph{sampling-without-replacement} (SWOR) eligibility criteria inspect a clients' Operational Stats (OpStats) database, a log of a client's previous task contributions which is kept by the Federated Compute client library. SWOR eligibility criteria can be used to ensure that a client does not contribute data more than once per task-defined period (e.g. no more than once every 24 hours, or only once in the lifetime of a task), which is a useful primitive for providing DP guarantees (see \cref{sec:differentially_private_heavy_hitters}).

\subsubsection{Ledger Attestation Verification and Uploading Encrypted Results}
\label{sec:ledger_attestation}
Once clients have run a summarization task and produced a result, the server provides clients with the ledger's attestation evidence, and a public encryption key generated and signed by the ledger binary. Clients then perform two key checks:
\begin{enumerate}
\item Verifying ledger attestation evidence, ensuring that it is rooted in a hardware root of trust such as AMD SEV-SNP, and that the stage0 firmware, kernel, and application binary hashes are all valid and that the provided encryption key was produced by the attested application binary.
\item Validating the data access policy that was provided alongside the task assignment.
\end{enumerate}

Upon successful completion of these checks, the client encrypts its data using the encryption key and the HPKE-based process described earlier, and uploads the encrypted data to a server-specified URI, at which point the encrypted data can be further processed server-side with TEE-hosted processing, subject to the ledger limiting access according to the data access policy as described earlier. 

\subsubsection{Enabling External Verifiability of the Data Processing}
\label{sec:enabling_external_verifiability}

There are multiple ways in which a client can validate the ledger's attestation evidence and the provided access policy. A simple approach uses an allowlist of hashes that is provided to the client device out-of-band. If the received attestation evidence and acccess policy are both allowlisted, then the client locally logs (in today's system for Android, to logcat \cite{logcat}) an \emph{attestation verification record} containing the attestation evidence and data access policy.

By examining the attestation verification record from a device's logs, a user would be able to manually follow the same steps as the device did to verify the identity of the ledger binary that generated the public key and the corresponding ledger implementation that protects access to the corresponding decryption key. The manually verified evidence can also prove that the ledger is running in a TEE protected by a hardware root-of-trust like AMD SEV-SNP. And by inspecting the data access policy, the full set of data processing steps and their corresponding implementations can be determined.

This simple approach requires gathering attestation verification records from devices one at a time. Its limitation is the infeasibility of instrumenting all devices that could possibly contribute data, which might (in the worst case) be necessary to discover the full set of access policies and binary endorsements served to and accepted by devices. In \cref{sec:enhanced_external_verifiability} we describe a more scalable approach to validating ledger attestation evidence and access policies based on a public transparency log.

\subsection{Data Processing}
\label{sec:data_processing}
When clients upload data, the encrypted data is written to storage, but as described in \ref{sec:ledger}, the data can only be decrypted with an in-memory key held in the ledger. This allows for processing of the data asynchronously from the upload step, and that decoupling offers simplicity and flexibility benefits over systems such as \cite{bonawitz2019towards}.

Theoretically, the entirety of the data might be processed in a single TEE; however, for the scale of workloads we aim to support, this would be infeasible in terms of compute and memory, and suggests the use of a graph of TEEs for parallel and pipelined processing of data. Our architecture distinguishes between \emph{orchestration}, which only ever operates on encrypted data and thus can run on hardware without confidential computing protections enabled, and \emph{data transformations}, which require decrypting the data and thus must run in TEEs. We define a data processing workload as consisting of untrusted infrastructure orchestrating a graph of data transformations that run in TEEs. For many applications (as we will see in the next section), this provides a sufficient degree of verifiability while also allowing flexibility in data processing patterns and infrastructure.\footnote{For example, central DP guarantees often require global consistency with respect to the ordering or grouping of incoming data. A user-level DP algorithm might, for instance, require that all of one user's data be processed on a single node for joint clipping. Or a DP learning algorithm might require that gradients in later stages are applied in an order pre-determined by an early stage. In these cases, early data processing steps must communicate global consistency requirements to later steps, which can assert that the untrusted orchestrator provides data in the expected way. Thus, a malicious orchestrator would be able to slow or halt progress (a processing step's assertions on data consistency fail) but would not be able to compromise the data processing pipelines' anonymization guarantees beyond what they might learn about ordering.}

\subsubsection{Orchestration}
\label{sec:orchestration}
The separation between untrusted orchestration and trusted data transformation allows us to leverage existing systems that are optimized for executing particular workloads. For example, for batch and streaming analytics workloads, orchestration can be performed by existing, widely used frameworks for writing and running map-reduce style pipelines, like Apache Beam \cite{apacheBeam}.

The orchestration layer starts up the TEEs that will execute the data processing. TEEs may have a high startup overhead due to the potentially expensive operations of VM memory validation and attestation evidence generation, which the TEE infrastructure must perform prior to being able to execute workload-specific code in a trustworthy manner. It then distributes the workload across the available TEEs to execute the desired transformations by passing configuration and encrypted data to the TEEs. Finally, the orchestration layer must handle TEE failures in a manner acceptable to the workload and data usage policies.

\subsubsection{Trusted Applications}
\label{sec:trusted_applications}
As is the case for orchestration, the data transformations all have different implementations depending on the workload, but the TEEs that execute the data transformations all have common responsibilities. Data transformations are implemented by a trusted application running in a TEE, which is launched by an untrusted companion application. On startup, a communication channel is established between these two applications, which is the only way for the trusted application to perform I/O. Given that communication between the trusted application running the data transformation and the ledger is mediated by the untrusted companion application, the protocol must be resistant to delayed, inserted, and replayed messages.

After startup, the untrusted companion application provides the trusted application with any workload-specific configuration. The trusted application then generates a keypair to be used for decryption, and signs the public key and privacy-critical configuration properties with a key rooted in the attestation. The untrusted companion application can then provide the signed public key and configuration properties and the attestation of the trusted application to the ledger in order to request access to data provided by the orchestration layer. The details of how communications between the ledger and the trusted applications are secured cryptographically are described in more detail in \cref{sec:ledger}.

When the trusted application receives data to be processed that has been authorized by the ledger, it decrypts the data using the private key that it generated at configuration time, then passes the decrypted data to the workload-specific data transformation code. Nonces uniquely identify authorizations from the ledger, thus preventing replaying of messages between ledger and the trusted application (e.g. efforts to process the same data twice).

Execution of the transformation over data must not be vulnerable to rollback or forking attacks that could compromise privacy. We use hardware that aims to protect the integrity of VM memory against a malicious hypervisor \cite{amdSEVSNP}. However, if the TEE crashes before the data transformation completes, any in-memory state is lost. For this reason we consider data transformation TEEs to be ephemeral. Fault tolerance is not a property of the data transformations themselves, but data transformations can be retried by the orchestration layer in order to provide fault tolerance at a higher level of the stack; for further discussion, see \cref{sec:robustness_and_scaling_of_stateful_components}.

Data processing steps may either produce encrypted, intermediate results intended for further processing in TEEs, or---for terminal steps in a data processing pipeline---they may attempt to release unencrypted results such as statistics, model weights etc. to the untrusted environment. For any output that is still considered private, the trusted application encrypts the output following the cryptographic protocol with the ledger, so that the output is subject to the usage constraints that are encoded in the data access policy. Only then is the encrypted output be passed outside of the TEE boundary to the untrusted companion application in order to be used in further processing steps. For terminal steps in the pipeline, the trusted application only releases unencrypted outputs that have been produced by a privacy preserving process, subject to the guarantees outlined in the data access policy (recall that the trusted application is part of the chain of trust, given that it has attested to the ledger and the ledger has verified that its attestation and configuration properties match the data access policy).

As in the case of orchestration, data transformation code can benefit from using libraries optimized for running particular workloads, such as SQL engines for analytics, and TensorFlow or XLA for training ML models. However, since the transformations operate within TEEs on private data, the code implementing transformations should be subject to increased scrutiny, as claims that the data is handled in a privacy-preserving manner are only as strong as the code that implements those privacy protections. The use of existing libraries within data transformation TEEs has the benefit of enabling more advanced data processing patterns, but comes with the drawback of increasing the size of the TCB and thus the code that must be verified in order to accept that the TEE is performing a given operation.

\subsubsection{Proprietary Functions and User-Level DP}
\label{sec:proprietary_functions}
Open-sourcing of code is the foundation of external verifiability, so a particular challenge to proving that data is handled in a privacy-preserving manner arises for workloads that have a requirement to run some proprietary code over data. Open-sourcing of all data transformation code enables making privacy guarantees that are based not only on formal DP properties, but also on empirical observations of the privacy loss of the given workload \cite{andrew2023oneshot} against weaker-than-worst-case adversaries. Open-sourcing the entire workload also provides the most flexibility in the processing pattern while still guaranteeing privacy properties. Despite the advantages to fully open-sourced data processing workloads, in order to enable broad applicability of our privacy-preserving infrastructure we must contend with the fact that many workloads contain proprietary portions, such as a proprietary query or model architecture. In these cases, a requirement to open-source the entirety of the data transformation code could prevent adoption of the privacy protections for user data that our system offers.

\begin{figure}
    \centering
    \includegraphics[width=0.7\textwidth]{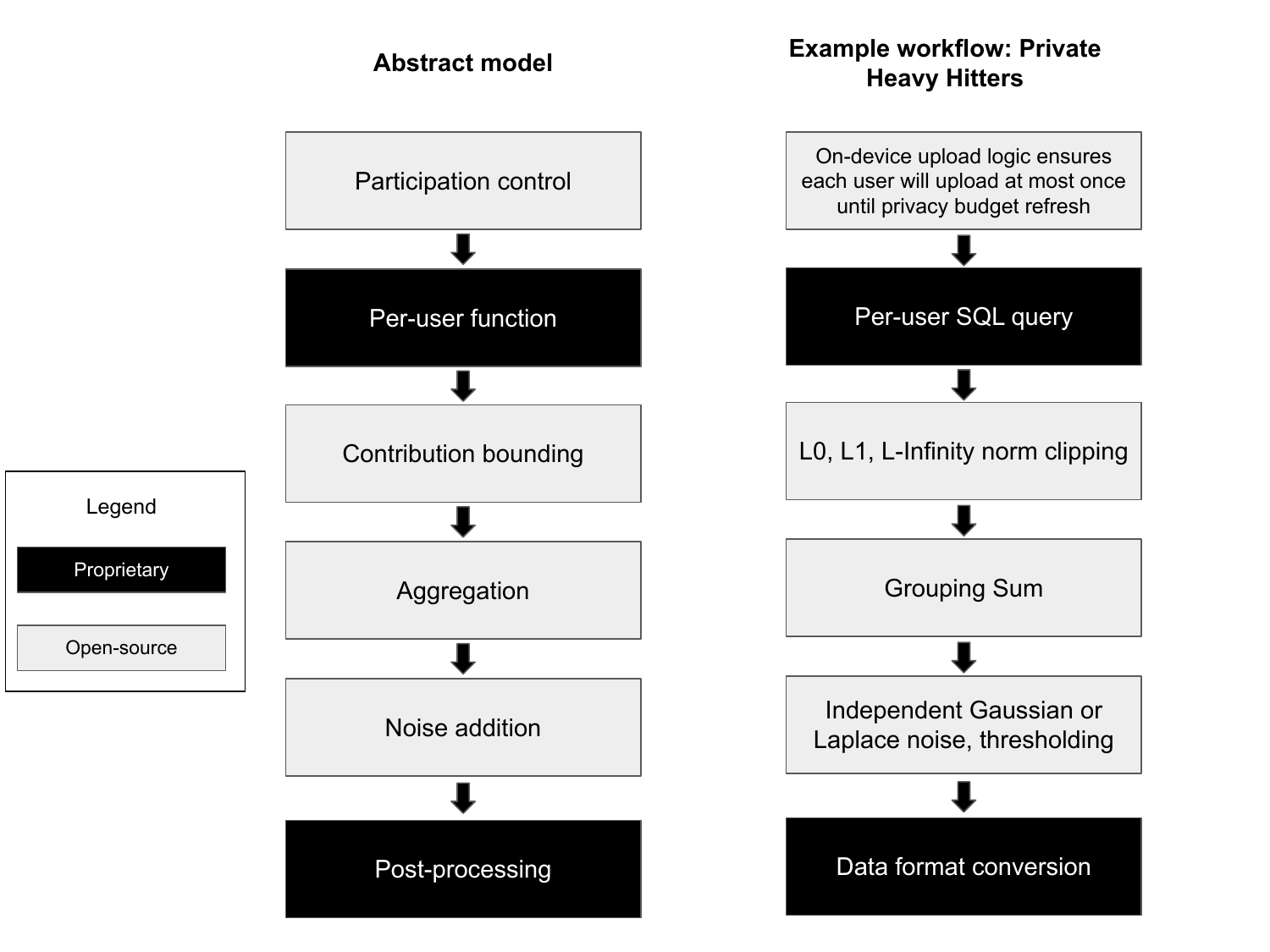}
    \caption{Ordering of DP operations relative to proprietary code and example Private Heavy Hitters application which will be discussed in \cref{sec:differentially_private_heavy_hitters}.}
    \label{fig:dp_operation_ordering}
\end{figure}

To contend with this challenge, we consider the proprietary portions of the workload to be untrusted functions that might run over user data, and reason about how we might limit the power of such functions to influence the privacy of the output of the overall data processing workload. We observe that by the definition of differential privacy \cite{dwork2006dp}, user-level differential privacy bounds hold regardless of the contents of the data from each user. Thus, even if an untrusted function independently modifies each user's data before the DP algorithm is carried out, we can still make formal guarantees about the amount of information that can be learned about a user from an output artifact. What this means in practice is that if our data processing workloads ensure that a given user's data can be modified by proprietary code only before executing a DP algorithm on the modified data (such an algorithm may include DP contribution bounding, aggregation, noising, etc.) we can still make a verifiable claim that the value of the DP $\varepsilon$ controls the adversary's ability to distinguish between a dataset containing a given user's data and a dataset without that user. Figure \ref{fig:dp_operation_ordering} shows the sequencing of DP operations relative to proprietary code that is required to preserve the desired properties. 

This observation only holds if the untrusted function has limited power: the function must only execute on one user's data at a time (optionally with other already-DP-protected inputs derived from other user's data, like the current model iterate) and produce a single output for that user. Consider an untrusted function that rather than taking an input from one user and producing one output, takes inputs from N users and produces N outputs. An adversarial version of this function could produce N outputs that all correspond to one of the input users, breaking the DP algorithm's assumption that each input corresponds to a different user. The same attack could be carried out if the untrusted function has the ability to save state between executions on inputs from different users. Thus, when writing trusted applications that execute proprietary data transformations, the externally verifiable open-source code must ensure that state is cleared between executions of the proprietary code.

This data processing pattern of per-user execution followed by aggregation is similar in structure to traditional federated algorithms. For this reason, we expect many data processing workloads will continue to use a federated processing pattern, even for workloads where the bulk of data processing logic might execute on the server-side within TEEs.

\subsubsection{Aggregation Cores}
\label{sec:aggregation_cores}
Given the expected prevalence of workloads where aggregation with DP is critical to the overall privacy guarantees, we believe it is important to build reusable and externally verifiable primitives for aggregation with DP. For this reason we are developing a library of \emph{aggregation cores}, which implement aggregation over arrays of data and DP operations on those arrays. Aggregation cores are designed to be performant so that they can be used for aggregating over billions of inputs or aggregating inputs containing multiple gigabytes of data. An explicit non-goal of the aggregation cores is flexibility. Unlike libraries like TensorFlow \cite{tensorflow2015-whitepaper} or SQLite which are designed for expressibility, aggregation cores will support only a minimal set of aggregation primitives with few configuration parameters, to facilitate reasoning about the properties of the privacy-critical DP aggregation step.

\subsection{Enhanced External Verifiability}
\label{sec:enhanced_external_verifiability}

In \cref{sec:enabling_external_verifiability} we described a simple allowlist-based mechanism through which client devices can validate and log ledger attestation evidence as well as the data access policies they are served, enabling a moderate level of verifiability for anyone instrumenting the device. We also established an important downside to this approach: that it doesn't allow one to determine the full set of ledger binaries or access policies that any given device, not just a specific instrumented device, might accept.

This section describes a complementary mechanism which achieves an enhanced level of external transparency, by enabling anyone to determine the \emph{complete} set of data processing steps that could apply to the data uploaded by \emph{any} device, without having to instrument any devices at all. The approach augments the architecture described in the previous sections with \emph{cryptographic endorsements} of binaries and access policies, which are then published to a \emph{transparency log}.

\paragraph{Ledger and access policy endorsements}
TEE binary endorsements---cryptographic signatures of the binary's digest---are used to indicate to a client device that a binary has been signed by the service provider to indicate its correct operation. By endorsing binaries for each layer of the TEE-hosted ledger (firmware, kernel, application, etc.), devices can verify that the service provider has claimed that uploads' decryption keys will be released only in accordance with those uploads' access policies. And of course, this claim can be verified by anyone who examines (and reproducibly builds) the ledger binaries resulting in the same endorsed binary digest.

Access policy endorsements---cryptographic signatures over an access policy's digest---are used to indicate to a client device that the policy has been signed by the service provider as appropriately privacy preserving. For example, an access policy endorsement could be used to indicate that only differentially private aggregates can be released to the service provider, a claim that can, in turn, be verified by examining (and reproducibly building) the binaries as associated with the processing steps referenced by the access policy.

Note that this architecture allows client devices to require ledger binaries or access policies to be endorsed by \textit{multiple} signing parties. A client application that requires multiple endorsements could, for example, require that the ledger binary or processing steps be vetted by an independent auditor (not the platform software authors). While today's production system does not currently use such configurations, we see this as promising area of future work.

\paragraph{Transparency logs}
Transparency logs are tamper-evident structures to which data is published in a non-revertible, append-only manner.
Violations of this append-only property are publicly detectable.
Transparency logs can also generate for each logged entry an \emph{inclusion proof}, or a cryptographic signature made by the transparency log operator that the entry has been included in the public log.
Inclusion proofs allow devices to verify signed items are transparent without interacting with the transparency service, as long as one trusts the log operator's handling of the inclusion proof signing key.

To enable public inspection of our ledger binaries and access policies, we publish the ledger binary and access policy endorsements described above to a transparency log. We use the Rekor \cite{rekor} transparency log service for this purpose, part of the SigStore project and operated by an independent non-profit organization.
The transparency log operator can thus be considered an adversary that is separate and non-overlapping with the other adversaries in our threat model (see \cref{sec:threat_model}).

Before accepting ledger attestation evidence and access policies through the endorsement verification process, client devices can also validate that endorsements also have a corresponding transparency log inclusion proof. In this way, external parties can monitor the transparency log service to determine the full of set of ledger binaries and data access policies that a client device could ever accept. Doing so requires only monitoring the log for new entries created by an endorsing key of interest.

Thus, external parties can determine all possible code paths through which uploaded client data might be processed. By inspecting the provenance (open-source code and build instructions) for all relevant TEE-hosted binaries (ledger and data processing steps) and access policies, anyone can follow along and validate or falsify the privacy properties of data processed in the system.

\subsection{OSS Repositories}
\label{sec:oss_repositories}
\paragraph{TEE-hosted kernel binaries and container images} The Project Oak repository \cite{oakOnGithub} contains the source code for container images and kernels used to run TEE-hosted application binaries, as well as instructions for mapping a kernel-layer or container-layer binary hash in the attestation evidence to a specific kernel version or container image.

\paragraph{TEE-hosted application binaries} \!\!The Confidential Federated Compute \cite{cfcOnGithub} repository contains the source code for the ledger binary and a number of data processing pipeline binaries, documentation of their implementations, and instructions for mapping binary digests (logged in an attestation verification record produced by a client, or endorsed in transparency log entries) to the corresponding reproducibly buildable binaries and their source code.

\paragraph{Client-side code} The code that runs on end-users devices and performs the verification of attestation evidence, logging of externally inspectable attestation verification records, and encryption of client payloads can be found in the Federated Compute Platform \cite{fcpOnGithub} repository.

\paragraph{Attestation verification library} The Oak Attestation Verification library \cite{oakOnGithub} is used by both the client-side code and the ledger binaries to verify the attestation evidence for TEE-hosted applications. The client-side code verifies the TEE-hosted ledger binary, while the ledger binary verifies other TEE-hosted data processing binaries as allowed by the data access policy.

\paragraph{Scalable multi-TEE services} To prevent state from being lost when the ledger TEE goes down, its implementation in the Confidential Federated Compute repo leverages the utilities for multi-TEE distributed services implemented in the Trusted Computations Platform \cite{tcpOnGithub} repository.

\section{Applications}
\label{sec:applications}
This section aims to illustrate the use of the platform with two common types of federated computations---open set histograms, and on-device learning---followed by a brief outlook on extended usage patterns.

\subsection{Differentially Private Heavy Hitters (PHH)}
\label{sec:differentially_private_heavy_hitters}
An initial use case we are exploring for the architecture described above is the privacy-preserving discovery of heavy hitters \cite{cormode2003finding,zhu2020federated}, more precisely open set histograms with differential privacy guarantees\footnote{For this application, we are considering device-level DP with a ``replace-one'' adjacency notion where one device's data can be changed arbitrarily.}. Devices maintain counts of items (each item being identified by a unique key) and our objective is to compute the aggregate counts across devices with a differential privacy guarantee. 

\noindent To realize this on the platform, data scientists define two tasks, which may be expressed in SQL: 
\begin{description}
\item [client summarization query] The client summarization task is responsible for selecting data from an on-device database and uploading the device's key/count pairs to the intermediate storage location as an encrypted blob.
\item [server-side workload] The server-side workload specifies TEE-hosted transformations that apply a differentially private aggregation of the counts. Data scientists can use an extension of GoogleSQL's differential privacy SQL options syntax to define a PHH query with specific DP parameters. This gets compiled into a protobuf configuration message consumable by a PHH-specific Aggregation Core implementation.
\end{description}

\noindent Listing~\ref{tab:serversql} shows an example server-side workload definition for a query that generates two histograms: one for the number of items purchased on weekdays and one for the number of items purchased on weekends (both keyed by color and food). A PHH COUNT query can be conveniently expressed via a more general SUM primitive, which also allows handling the case where devices can contribute a count greater than one per item. Note how the cross-device aggregation is a sum of counts, not a count; more generally, our system supports differentially private summation of numerical values associated with keys.

\begin{center}
\begin{tabular}{c} 
 \begin{lstlisting}[caption={Server-side workload query example}\label{tab:serversql},basicstyle=\small]
SELECT WITH DIFFERENTIAL_PRIVACY OPTIONS
      (epsilon=1, delta=1e-8, max_groups_contributed=2)
  color, food,
  SUM(num_purchased_weekdays) @{L_inf = 3} AS total_num_purchased_weekdays,
  SUM(num_purchased_weekends) @{L_inf = 4.5, L_1 = 8, L_2 = 6} 
      AS total_num_purchased_weekends,
FROM 
  -- table representing all uploaded data with columns
  -- color, food, num_purchased_weekdays, num_purchased_weekends 
  uploaded_device_data       
GROUP BY
  color, food;
 \end{lstlisting}
\end{tabular}
\end{center}

Note that the server-side workload definition specifies a privacy budget in terms of epsilon and delta ($\varepsilon$, $\delta$), which is the maximum privacy loss that the server-side workload will incur. The platform will match the query against the required data access policies associated with the user data being accessed (not shown). Before releasing any uploaded client data to the server-side processing pipeline, the platform confirms that the server-side workload definition conforms with the access policy. In this example, a data access policy requiring  $(\varepsilon^*, \delta^*)$-DP or stronger would permit the query from Listing~\ref{tab:serversql} to run as long as  $1 \le \varepsilon^*$ and $10^{-8} \le \delta^*$. The access policy can also limit how often the data is used, for example specifying that any uploaded client data will be processed at most once by a TEE-hosted binary containing PHH Aggregation Cores.

In addition to the privacy budget, there are four additional DP parameters that the PHH Aggregation Core implementation uses and that data scientists are able to set using SQL hints. These clipping parameters define the differential privacy sensitivity. They instruct the aggregation core to clip per-device data to these bounds and are also used to determine the differential privacy noise setting and the key release threshold. These parameters therefore primarily affect the utility of the resulting data. They are:
\begin{itemize}
\item \texttt{max\_groups\_contributed} ($\ell_0$): Each device is only allowed to contribute to a maximum of $\ell_0$ keys.
\item \texttt{L\_inf} ($\ell_\infty$): Each device is only allowed to contribute a val below $\ell_\infty$ to a given key.
\item \texttt{L\_1} ($\ell_1$): Each device is only allowed to contribute a total val below $\ell_1$ across all keys.
\item \texttt{L\_2} ($\ell_2$): Each device is only allowed to contribute values for whom the root of the sum of squared values across all keys is below $\ell_2$.
\end{itemize}
For more details on how these parameters are applied to provide DP guarantees, see Appendix \ref{sec:phh_algorithm_details}.

The approach described above requires clipping parameters to be specified up front by the query author. Obtaining the best choice of parameters demands nontrivial knowledge about the distribution of the data: an overly small choice of $\ell_\infty$, for example, would introduce significant bias in each aggregate while overly large values would cause the DP noise to have overly large variance. Future work will explore a two-pass approach, where the first pass searches for clipping parameters without releasing a histogram\footnote{This can also be done by computing certain quantiles on quantities like ``number of items per user'' and using those to set the parameters.} and the second pass applies the query with the identified parameters. Depending on the use case and the data analyst's needs, the data access policy may require the computation (and release) of the parameters with DP.

The current algorithm operates in a batched-input-and-single-output setting, where user data for a specified time period, say a week, is uploaded only once via the use of on-device SWOR (see \cref{sec:client_side_architecture}), all user data is accessible at once, and only one differentially private report is required. In future applications we plan to support multiple uploads per-client and streaming releases, while continuing to ensure user-level differential privacy.

\subsection{Cross-Device Federated Learning}
\label{sec:on_device_federated_learning}
Another use case of interest is cross-device federated learning \cite{mcmahan17fedavg}, which has been the focus of Google's existing system  \cite{bonawitz2019towards}. A key property of the architecture proposed here is the decoupling of device side processing (computing summary statistics, gradients, uploading results) from subsequent server side processing (e.g. aggregation over large amounts of data or longer periods of time). This is a natural and scalable solution for analytics use cases where client summaries are independent of each other (i.e. one client's results do not depend on other clients); learning use cases on the other hand tend to be iterative in nature, requiring feeding back results such as the latest model from the aggregation pipeline to devices running the gradient computation.

There are roughly two strategies for how to orchestrate learning computations that have such a feedback loop. One is \emph{asynchronous FL}, notably the TEE based approach pioneered by Papaya \cite{huba2021papaya}---the aggregation pipeline emits results (new models) that are immediately served to newly connecting clients, and client contributions are (subject to staleness checks) immediately incorporated into the currently running aggregation. This processing strategy is straightforward to support in our proposed architecture.

The other strategy is \emph{synchronous FL}---running a sequence of ``rounds,'' each operating on a closed set of devices (a \textit{cohort})---which is the approach used in the original federated learning paper \cite{mcmahan17fedavg} and system architecture \cite{bonawitz2019towards}. This approach may increase system complexity \cite{huba2021papaya,kairouz2019advances}, but synchronous FL has been highly successful in the academic literature as well as existing use cases at Google, thus motivating continued support by our new architecture. 

\begin{figure}
    \centering
    \includegraphics[width=0.85\textwidth]{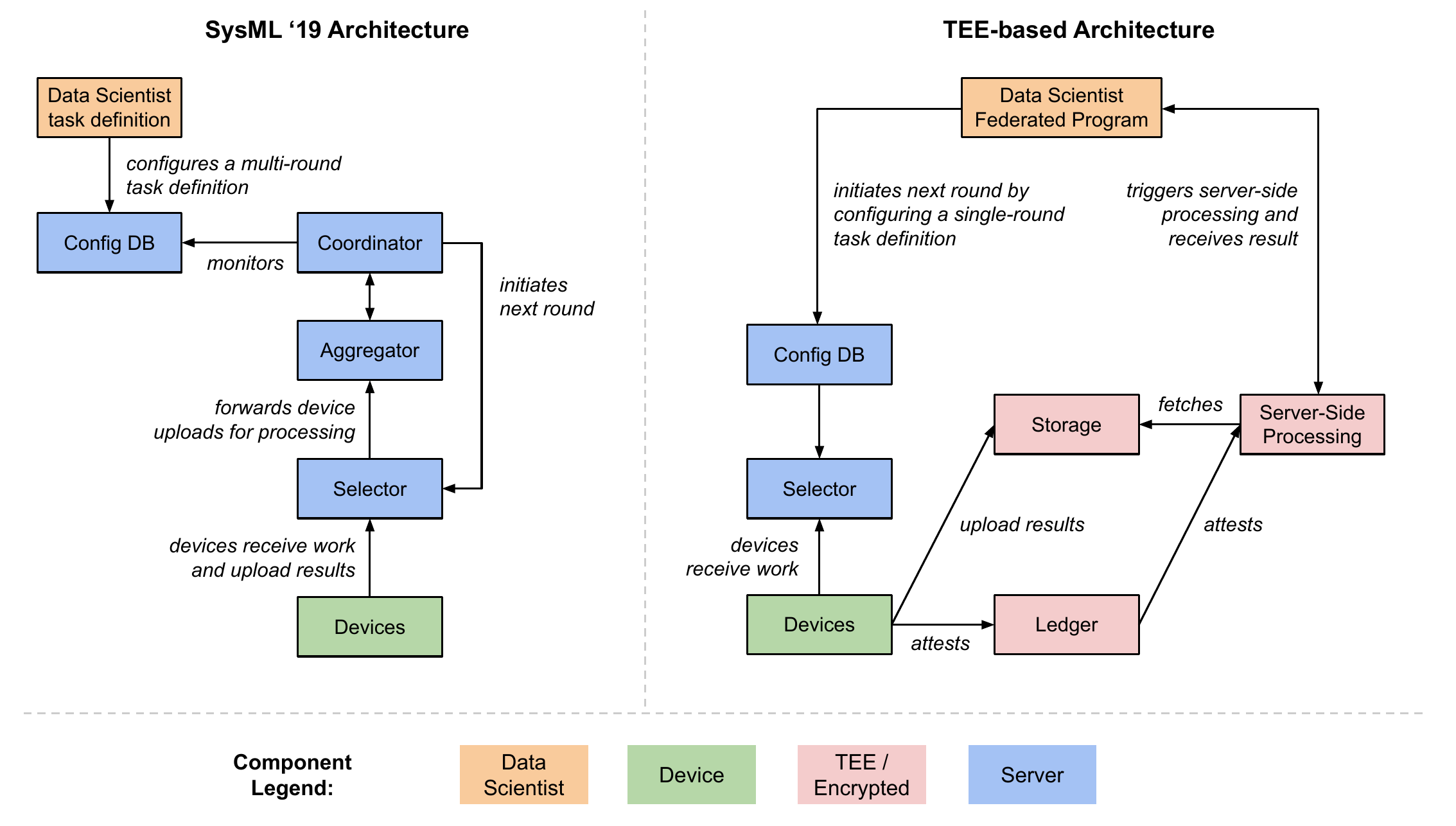}
    \caption{The proposed architecture uses a Federated Program to orchestrate round progression instead of a Coordinator.}
    \label{fig:round_orchestration}
\end{figure}

While on-device computations do not change, round progression for synchronous FL workloads must be orchestrated differently in the new architecture, as illustrated in figure \ref{fig:round_orchestration}.

In the original system, rounds are orchestrated by the Coordinator object (see Section 4.2 and Figure 3 in \cite{bonawitz2019towards} for more detail). In contrast, in our new TEE-based architecture, a novel \emph{Federated Program} \cite{tffFPOnGithub} component is the orchestration layer responsible for deciding when to request work from a new round of clients and when to run server-processing logic over a set of encrypted client uploads. When sending task configuration updates (e.g. start/stop/slow) to the Config DB, the Federated Program can incorporate custom signals, such as monitoring the rate and number of uploads received by the intermediate storage location. The Federated Program can also take a variety of approaches with running the server-side processing logic. Processing a fixed set of client uploads that are all associated with the current round is simplest, however incorporating client uploads in a streaming manner or incorporating straggler client uploads that were associated with previous rounds are possibilities too. In general, aiming for high client resource utility (i.e. not wasting work performed by clients) via the Federated Program orchestration layer is preferable.

In both orchestration approaches shown above, the per-client gradients are computed on-device, meaning an identical amount of data is being uploaded from the client to the server and existing data minimization practices are upheld. However, the two approaches differ in the lifetime and storage properties of pre-aggregated data on the server. In the previous system, client uploads are immediately aggregated in-memory, whereas in the TEE based architecture, ledger-encrypted client uploads are stored in an intermediate location prior to being accessed by a server-side processing pipeline launched by the Federated Program. Under our threat model described earlier, storing ledger-encrypted pre-aggregates in e.g. a ramdisk or persistent storage prior to aggregation does not meaningfully increase the attack surface compared to immediate in-memory aggregation; when paired with the externally verifiable DP aggregation, we believe the architecture provides stronger privacy guarantees.

\subsection{Confidential federated learning leveraging server-side processing}
\label{sec:confidential_federated_learning}
Cross-device federated learning can be expressed as a confidential federated computation, offering complete transparency into the server-side steps involved in combining per-device model updates. The strong transparency guarantees made by confidential federated computations let federated learning go a step further: model gradients can themselves be computed server-side where appropriate, enabling training and evaluation of much larger models. The relation of this idea to traditional cross-device federated learning is explored in more depth in \cite{daly2024federated}.

\noindent To represent sever-augmented federated learning as a confidential federated computation, data scientists define two tasks:
\begin{description}
\item [client summarization task] responsible for selecting data from an on-device database, which will be encrypted with a ledger-managed public key before upload to a temporary data store.
\item [server-side workload] as represented by a script (a \emph{Federated Program}), describing the multi-round training process and/or evaluation process to run over the uploaded data.
\end{description}

When devices upload, their messages are cryptographically bound to a publicly-inspectable access policy naming the open source federated program and its corresponding processing steps. A canonical program will process and aggregate many users' data, for example by using a differentially private aggregation learning algorithm. Some of the program's arguments may not be public---for example, proprietary model definitions or weights might be arguments bound in the host environment at runtime---which need not affect the external verifiability of the privacy guarantees if the proprietary components affect only the ``per-user function'' and ``post-processing'' operations shown in Figure \ref{fig:dp_operation_ordering} and discussed earlier in \cref{sec:proprietary_functions} (although efficacy of side-channel attacks likely increases).

The federated program encodes its own logic for selecting data for round-based processes, as this is generally required for proper DP accounting. The access policy specifies the maximum number of times the program may be executed using a specific piece of data as an input, limiting the risk of a program that is run mulitple times to triangulate away added noise. 

Execution of the server-side workload uses an untrusted orchestration layer to configure a \emph{program executor TEE} to host the federated program and to inject any proprietary models or data. First, an initial configuration of the program generates TEE attestation evidence used throughout the federated program execution to obtain the ledger's approval for access to data with a matching access policy. Model weights, initial checkpoints, and any other (potentially proprietary) data needed by the program are provided to the TEE and incorporated into the loaded federated program via a separate side-channel configuration mechanism after attestation. The expressiveness of these side-channel configuration inputs is heavily constrained toward preventing a malicious orchestrator from overwriting a previously loaded federated program.

To enable efficient execution, the program executor TEE may delegate execution to a collection of attested worker TEEs, e.g. to compute gradients across many uploads. The chain of trust is extended from the program executor TEE to each worker TEE via the noise protocol \cite{perrin2018noise}, wherein the worker TEE provides its attestation evidence to the program executor TEE and a secure communication channel between the two TEEs is established. The untrusted orchestrator forwards messages containing end-to-end encrypted data between the program executor TEE and its worker TEEs. Only the program executor TEE has the ability to release unencrypted results into untrusted space, as denoted by an explicit step in the federated program. Portions of the federated program that are included in the access policy describe the points during execution at which unencrypted data may be released.

In comparison to cross-device federated learning, server-side confidential federated learning enables training of much larger models. Far less work is performed on-device, so device resource constraints (memory, bandwidth, and compute) no longer limit the effective size of the model. Server-side TEEs may also enable the use of new algorithms, including those that join with server-side data, or make use of other data processing patterns that would be impossible in cross-device federated learning.

\subsection{Future Applications}
\label{sec:future_applications}
The described architecture enables a multitude of other applications we are yet to explore:
\begin{itemize}
\item Cross-silo federated learning could be drastically simplified by processing data centrally in TEEs, instead of a training infrastructure that is distributed across independent and possibly mutually distrusting organizations.
 \item Confidential federated learning can leverage more server-side compute resources than traditional cross-device federated learning, and open the possibility of much larger models, novel algorithms (such as joins with server-side data) and efficient hyperparameter search (DP and learning parameters), as outlined in \cref{sec:confidential_federated_learning}; .
\item Today's differentially private learning algorithms must be designed to take into account that intermediate model updates are ``public'' (visible to end user devices, where the next set of gradients is computed); a TEE learning pipeline on the other hand will emit only the final result to the untrusted environment and may therefore provide significantly better empirical and formal differential privacy guarantees.
\end{itemize}

\section{Open Problems and Future Work}
\label{sec:open_problems}
\subsection{Scalability and Robustness Challenges}
\label{sec:scalability_and_robustness}
For applications that process limited amounts of data and do not need to provide high availability, executing the entire trusted logic---key management, data processing, aggregation---within one physical TEE may be a compelling option due to its architectural simplicity. For most real world applications however, that approach
\begin{enumerate}
\item Does not scale in space and time to larger workloads.
\item Does not provide sufficient fault tolerance.
\item Does not provide composability / flexibility.
\end{enumerate}

The majority of use cases and scales thus require distributed systems of TEEs. We have already seen how chains-of-trust can enable parallel and pipelined processing; however, various performance and security concerns remain, and we will briefly touch on them in the following sections.

\subsection{Per-TEE Overhead and Limits}
\label{sec:tee_overhead}
Besides the expected overhead of memory encryption, applications running confidential VMs based on TEE technology incur overhead (relative to traditional VMs) due to copying input data from shared to private memory, followed by decryption, as well as the cost of protections for memory integrity and isolation, e.g. the RMP checks in AMD SEV-SNP.

Our target applications require that VMs ingest significant amounts of data. This results in I/O intensive workloads, which are the type of workload expected to show more delays \cite{intelTDXPerformanceConsiderations}. This is mostly due to repeated copying from shared to private memory, and the overhead of the corresponding transitions between trusted and untrusted space. Our preliminary measurements match this expectation, with overhead in CPU intensive workloads being significantly smaller than the one observed in I/O intensive workloads. Overcoming this limitation involves a careful configuration of VM devices to maximize CPU utilization, and batching on the untrusted side.

CPUs in and by themselves may pose another scalability concern---modern machine learning is relying heavily on accelerators such as GPUs or TPUs, but availability of Trusted Computing support on these processors is very limited at this time, and further infrastructure investment to integrate them securely with CPU TEEs will be needed.

\subsection{Robustness and Scaling of Stateful Components}
\label{sec:robustness_and_scaling_of_stateful_components}
The state stored in the ledger includes the history of data usage, which if rolled back or forked could lead to surpassing the privacy budget for that data. In order to guarantee privacy in the face of a malicious and colluding platform software operator and data center storage operator, while also guaranteeing liveness of the state in the face of TEE failures that would occur during normal operation of a system,  this state has to be protected from rollback, forking and replay attacks. Specialized hardware might be an attractive option, however it lacks desirable properties, namely external verifiability and reproducibility, ease of hardware provisioning and performance. Therefore our solution opted into using the TEEs available in commodity hardware to preserve integrity and confidentiality during computation through the utilization of techniques such as memory isolation, encryption, and remote attestation. TEE serves as a fundamental building block for privacy mechanisms that can be externally verified. However, using it to power privacy focused computations at Google scale requires careful balance of security, scalability and resource costs, that we are going to briefly describe (with the full details explored in an upcoming paper).

A physical TEE has inherent limitations of the memory and computation capabilities of the host machine. We argue that to address the scale of target workloads these limitations must be solved in conjunction with the mitigations of failures and active attacks (for example, an attempt to solve memory limitation via swapping into sealed storage creates new opportunities for state rollback attacks).

First, we start by addressing the fault tolerance and active attacks. Our focus on commodity hardware (instead of specialized hardware) leads us to the distributed systems solution. Specifically, we leverage physical TEE properties and hardened Raft consensus protocol to address fault tolerance and protect from rollback, forking and replay attacks. The Raft consensus protocol emerges as an optimal choice for essentially trusted replicated state machines (TRSM), owing to its emphasis on simplicity and comprehensibility.

Second, we need to address the challenge posed by substantial data volumes and processing costs. TRSM enables stateful fault tolerant confidential computations, however they are still limited by the host machine capabilities. Specifically it is not feasible to either hold or process all the state in a replicated state machine. At scale, cost factors escalate significantly, including memory demands for state storage, network bandwidth required for write replication and state snapshot distribution, computational requirements for data processing and privacy mechanisms, and the potential impact of failure recovery time. Therefore the replicated state machine must manage minimum required state and perform minimum processing, while offloading the storage to a non-TEE persistent blob store and the computation to non-replicated TEE nodes.

Third, we focus on the granularity of private state updates. Fine reads and writes incur high persistent blob storage overhead (lots of small blob reads and writes) as well as privacy overhead (remote attestation and encryption). The coarse reads and writes are crucial for the cost amortization. Therefore private data is organized and processed as coarse atomic units, thus reducing the state maintained by the replicated state machine.

See Appendices \ref{sec:related_work_on_rollback_protection} and \ref{sec:related_work_on_fault_tolerant_confidential_computations} for further pointers to related work in confidential rollback protection. Also see the Confidential Federated Compute \cite{cfcOnGithub} and Trusted Computations Platform \cite{tcpOnGithub} open source software repositories for evolving implementations of these ideas.

\subsection{Side Channel Attacks}
\label{sec:side_channel_attacks}
Offerings such as AMD SEV and Intel TDX aim at providing integrity and confidentiality even in the presence of an untrusted hypervisor and host system. Generally speaking, this kind of adversary tries to infer information from a given confidential workload/guest by manipulating the state and availability of shared resources handled by the host system, such as for example (cache) memory. A common source of attacks is side-channels, which traditionally include timing and memory access patterns, in addition to the more advanced speculative execution side channels. Both AMD SEV-SNP and Intel TDX mitigate side-channels by means of constant time cryptographic implementations, and data-independent access patterns. In both cases, the guest software is responsible for doing the same for sensitive operations. For reference, see \cite{intelSideChannelGuidelines}.

It is worth noting that the privacy guarantees of our platform crucially rely on secrecy of both (i) private cryptographic keys generated/held by a TEE, as well as (ii) noise samples and intermediate values in differentially private mechanisms such as the ones in the applications described above. Side-channel leakage might enable an adversarial hypervisor or host system to mount an attack to degrade the privacy guarantees of the system, by leaking information about (i) or (ii). While our implementation does not currently include application-level mitigations to architectural side channels, this is an area to be revisited in the future. Next, we describe some of the traditional side channels in the context of the target guarantees of our system, and potential mitigations.

\paragraph{Memory activity and ciphertext leakage} A common side channel leveraged by many attacks is the so-called page-fault controlled side channel. AMD SEV-SNP does not provide mitigations against tracking VM access patterns through page tables (see Table 1 in \cite{amdCiphertextVisibility}), and recommends that Guest code mitigates this leakage by means of implementations that avoid secret-dependent memory access patterns. In our setting, this corresponds to access patterns that might \emph{either} leak information about (i) or (ii) above. A mitigation considered in both the literature and real-world deployments of TEE technology \cite{signalOram,bogatov21epsolute} is Oblivious Random Access Memory (ORAM), an abstraction layer over memory that randomizes arbitrary sequences of access patterns to effectively make arbitrary code data-independent (regarding memory patterns). While ORAM has been used in practice, it has significant overhead. Both tweaking existing constructions to match production needs/requirements, e.g. when choosing parameters, and designing application specific data-oblivious algorithms, e.g. for the sort of uses discussed in \cref{sec:applications}, are interesting venues for further research to improve the system.

Additionally, a related side-channel in AMD SEV-SNP worth mentioning is \emph{ciphertext leakage}: AMD uses AES for memory encryption, with the XOR-Encrypt-XOR mode using the physical memory location as tweak values. This essentially results in a deterministic form of encryption, where an adversary with access to encrypted memory, e.g. a corrupted host, might observe whether a given access to a memory location changes the underlying plaintext. See \cite{amdCiphertextVisibility} for a detailed description and suggested mitigation strategies.

\paragraph{Timing and payload sizes} Runtime and payloads that depend on secret data are known to be exploitable to extract secret keys in cryptographic implementations. This sort of leakage might be also relevant in the context of differential privacy, as discussed in previous works \cite{bogatov21epsolute, haeberlen11dp}. An important observation is that in a system that aims to provide a DP-style guarantee, mitigations might be formalized in the same framework. This is the approach taken in \cite{bogatov21epsolute} to address payload size leakage.  Another recurrent operation that might be subject to side-channel analysis by an attacker is noise sampling. Even beyond side-channels,  implementation of DP mechanisms has subtleties: it has been shown implementations of floating point arithmetic as well as rounding schemes might result in unintended leakage \cite{desfontaines22dp}.

The space of side-channels is complex and evolving. As discussed in \cref{sec:threat_model}, we did not implement specific mitigations in the current version, but this is an area that we expect to evolve with feedback from the broader security and privacy community.

\subsection{Externally verifiable TTLs, Trusted Clocks, and Remote Deletion}
\label{sec:externally_verifiable_ttls}
As discussed in \cref{sec:ledger}, a limitation of Trusted Execution Environments is their lack of access to a trusted time source. The current version of our system mitigates this limitation through a few avenues. Firstly, we limit how data can be used through policies that rely on concrete usage count limitations rather than just TTLs. Secondly, as mentioned in \cref{sec:ledger}, clients run their own validation that the time reported by the ledger is close enough to the real time before uploading data. However, neither of these prevent encrypted data from being accessible for longer than was promised to the client by the HPKE key TTL, if the untrusted portions of the stack provide the ledger with out-of-date timestamps after client data has already been uploaded. As discussed in \cite{anwar19applications}, even if a TEE establishes secure communication with an NTP server, which should be trusted to provide the correct time, an attacker with control of the network or hypervisor can delay the response packets. Roughtime \cite{roughtime} aims to solve this by incorporating nonces into the protocol between client and NTP server, so that by virtue of the fact that the client-generated nonce is signed along with the response, the client knows the Roughtime server generated the response after the client sent the request. The client need not trust a single server, either, as the protocol supports contacting multiple servers to obtain cryptographic proof that a server is misrepresenting the time. In the future we would like to consider integrating the ledger with Roughtime or similar systems in order to provide more accurate guarantees about how long data remains accessible on the server. However, Roughtime remains at an early stage in which the number and availability of servers is limited. Additionally, given the interactive nature of the Roughtime protocol and the potentially high QPS to the ledger from data processing workloads, we would need to determine a strategy for limiting the load on the Roughtime servers while still maintaining a relatively accurate current timestamp.

Eventually, it may be desirable to offer client devices the capability to remotely delete data that was previously uploaded to the server, if the device owner decides the data should be deleted before the TTL expires. In some respects, this is simple to support with the protocols described so far, as the ledger can simply record that no remaining usage is allowed for the data. There are three main challenges that are left as future improvements:
\begin{enumerate}
\item It is desirable to verify that the deletion request is originating from the client that produced the data, which requires introducing client identity.
\item Remote deletion requires proof that the deletion was successfully processed by the same ledger instance that was entrusted with the original data---even if the HPKE key has already expired. This requires introducing ledger identity that outlives a single HPKE key.
\item Since remote deletion requires interactive communication between client devices and the ledger, the ledger must be made robust and scalable enough to serve traffic from client devices in addition to the traffic arising from data processing workloads.
\end{enumerate}

A related capability is allowing client devices to extend the lifetime of uploaded data beyond the original TTL. Combined with shorter TTLs, this capability would allow uploaded data to be deleted quickly unless clients actively sent keep-alive messages. Since the AEAD wrapping occurs independently from data encryption, this can be scalably implemented by re-wrapping the AEAD key with a newer HPKE key---but like remote deletion, it requires client identity to ensure that the requestor is authorized to extend the data TTL.

\subsection{Combining TEEs with SMPC}
\label{sec:combining_tees_with_smpc}
Secure Multi-Party Computation (SMPC) can be intuitively understood as a branch of cryptography that allows two or more parties to collectively simulate a trusted third party, through the use of distributed cryptographic protocols.  SMPC is often seen as an alternative answer to TEEs for the question of how to effectively distribute trust; so long as a (protocol-dependent) fraction of the parties participating in an SMPC computation behave honestly, the entire simulated trusted third party can be proven to behave as specified. An important aspect of the notion of security provided by SMPC protocols, is that no party (or valid coalition under the threat model) in a secure protocol execution has unilateral access to the raw data. This is because, in SMPC protocols, data is manipulated either under encryption or secret-shared, therefore avoiding a single point of failure.

While SMPC and TEEs are generally viewed as alternatives, we believe that it is a fruitful research direction to consider layering these two technologies together in order to achieve best-of-both-worlds defense-in-depth solutions. While the overhead of SMPC protocols that can withstand an active adversary is still significant (as opposed to semi-honest security), TEEs provide external verifiability that can strengthen the semi-honest security guarantee. For any operation of the system that can be implemented as an SMPC protocol over a small number of parties, we could implement each of those parties in its own TEE hardware, such that one would have to compromise multiple TEEs (in many cases, a majority) in order to compromise the SMPC protocol itself.  

In the simplest case, all the TEEs implementing the protocol could be identical and running in the same datacenter.  However, we can further improve resilience by using TEEs from multiple manufacturers (meaning that more than one hardware manufacturer/designer would need to be compromised to break security).  We can also run those TEEs in multiple datacenters, potentially operated by different operators, each with their own implementations of best practice security and resilience protocols against insider adversaries or coercion.

We believe that the ledger and the core aggregation routines would be particularly amenable to these defense-in-depth practices, drawing on experience with e.g. 2-party SMPC protocols such as Prio \cite{corrigan17prio}, and basic primitives amenable to efficient evaluation in MPC like distributed (threshold) key generation and signing.

\section{Conclusion}
\label{sec:conclusion}
We presented a system architecture for executing federated computations leveraging TEEs, accompanied by open-sourcing the relevant code paths running in TEEs and on end user devices. This architecture allows for improved DP-utility trade-offs, a more scalable system design, and robustness to Sybil attacks. To the best of our knowledge, this is the first time that the server side privacy properties of federated computations are externally verifiable. We are excited about the progress and opportunities this enables, and expect ongoing and future research, development and exchange with the academic community to further improve the robustness of the system.

\section{Acknowledgements}
Like most large scale system efforts, there are many more contributors than the authors of this paper. The following people have directly contributed to coordination, design, implementation and reviews: Tom Binder, Zachary Charles, Stanislav Chiknavaryan, Allie Culp, Sarah de Haas, Stefan Dierauf, Prem Eruvbetine, Zachary Garrett, Emily Glanz, Zoe Gong, Conrad Grobler, Steve He, Mira Holford, Wei Huang, Ulyana Kurylo, Artem Lagzdin, Katsiaryna Naliuka, Grace Ni, Ernesto Ocampo, Ivan Petrov, Juliette Pluto, Edo Roth, Andri Saar, Maya Spivak, Rakshita Tandon, Yu Xiao, Zheng Xu, Chunxiang Zheng.

\bibliographystyle{plainnat}
\begin{small}
\bibliography{references}
\end{small}

\appendix

\section{PHH Algorithm Details}
\label{sec:phh_algorithm_details}
As is standard for open domain DP histogram algorithms, we bound the contributions of clients, aggregate the bounded contributions, then add noise followed by thresholding. By controlling how much any single client can influence the aggregation, the noise we add will mask the signal from any single client. Variants of the algorithm we use are well-established in the academic literature (Algorithm 1 in \cite{bun16simultaneous}; Algorithm 1 in \cite{korolova09releasing}; \cite{googledp20calculations}). We summarize it below for the sake of completeness. 

\begin{center}
\begin{tabular}{ |c|c|c|c| } 
 \hline
 \multicolumn{2}{|c|}{Composite Key} & Value 1 (weight) & Value 2 (price) \\ 
 Key 1 (color) & Key 2 (food) & &  \\ 
 \hline
 green & eggs & $3.2$ & $2.99$ \\ 
 red & apple & $3$ & $3.99$ \\
 white & grape & $2$ & $3.49$ \\
 red & apple & $1$ & $2.99$ \\
 \hline
\end{tabular}
\end{center}

\begin{enumerate}
    \item For each client's table of data, such as the one above
    \begin{enumerate}
        \item Create the \emph{local histogram}, in effect a table with de-duplicated composite keys. In the above example, the local histogram would have three rows, one of which is a ``red apple'' row with weight $4 = 3+1$ and price $6.98 = 3.99+2.99$.
        \item Perform contribution bounding on the local histogram, using the parameters in the query:
        \begin{enumerate}
            \item \texttt{max\_groups\_contributed} limits how many rows are in the local histogram (i.e. how many distinct composite keys are impacted by one client), chosen at random; if the value is $2$, we would have a table with two rows. The contents of the surviving rows are untouched.
            \item When given as a type hint to $\sum{\texttt{value}_j}$, $\ell_\infty$ bounds the magnitude of a single value in the $j$-th column of the local histogram. In the Weight column of the above example, the weight of $3.2$ exceeds the $\ell_\infty$ bound of $3$ in the query, so it would be replaced by $3$. Meanwhile in the Price column, the aggregate $6.98$ would be replaced by $4.5$. All other data would be untouched.
            \item $\ell_1$ and $\ell_2$ are parameters that can be given to bound client data further and thereby reduce the scale of noise, but DP can be achieved without them.
        \end{enumerate}
    \end{enumerate}
    \item Perform a GROUP-BY SUM on the local histograms to produce a \emph{cross-client histogram} $H$, where the grouping is by the keys and the sum is over the values.
    \item Add noise to every aggregated value in $H$:
    \begin{enumerate}
        \item The scale of noise grows with $\texttt{max\_groups\_contributed} \cdot \ell_\infty$, which captures the total influence of a single client's local histogram. If the data scientist provides an $\ell_1$ value that is less than $\texttt{max\_groups\_contributed} \cdot \ell_\infty$, it is used instead. The same holds for $\ell_2$.
        \item If there are $p$ sets of values being aggregated—columns in $H$ that do not hold keys—the scale of noise grows with ($p / \varepsilon$). In the running example, $p$ = 2 because the only value columns are price and weight.
        \item The distribution of noise is either Laplace or Gaussian, whichever will result in lower variance for the given parameters.
    \end{enumerate}
    \item Drop a row from $H$ if any of its noisy values have magnitude less than $t_j$, where $t_j$ is a threshold computed from the algorithm's parameters in column $j$ ($\varepsilon$, $\delta$, \texttt{max\_groups\_contributed}, $\ell_\infty$, $\ell_1$, $\ell_2$). This step is necessary for DP because we must protect the set of composite keys as well as the sum of values: replacing one client's data can affect whether or not ``green eggs'' is present in the data, so the decision to include a row of data in the output should be noisy. 
\end{enumerate}

\section{Related Work on Rollback Protection}
\label{sec:related_work_on_rollback_protection}
Privacy mechanisms are stateful (tracking data or user privacy budget, storing privacy algorithm state and intermediate results, etc.). TEEs protect in-memory data through isolation, encryption and integrity mechanisms, however it doesn't provide fault tolerance (data is lost in case of failures or restarts). Many TEEs provide sealing capabilities to offload encrypted and signed state to disk that combined with monotonic hardware counters can be used to implement rollback protection. However this approach suffers from low performance, weak security and quick exhaustion of the hardware counters. Therefore a number of related works (including ours) followed a distributed systems approach, where a group of TEEs provide rollback protection. ROTE \cite{rote} provides monotonic counter service running in a group of TEEs and using the majority to perform reads and writes with conflict resolution. Nimble \cite{nimble} focuses on maintaining named hash chains (ledgers) using the majority of TEEs with witness co-signing. Unlike ROTE, Nimble provides a reconfiguration mechanism that is critical for production systems. The main difference from related work is our holistic focus on private data processing rather than rollback protection alone, including an end-to-end verifiable chain of trust.

\section{Related Work on Fault Tolerant Confidential Computations}
\label{sec:related_work_on_fault_tolerant_confidential_computations}
Confidential computations are typically long running and must be resilient to unexpected failures and planned maintenance. CCF \cite{russinovich2019ccf} and EngRAFT \cite{engraft} are using replicated state machines based on Raft consensus protocol running in a group of TEEs, both support reconfiguration. EngRAFT provides rollback protection for the Raft \cite{raft} state using a mechanism similar to ROTE. Avocado \cite{avocado} focuses on an in-memory key/value store running ABD \cite{lynch97robust} protocol in groups of TEEs. Avocado leverages key independence to achieve high performance (operations are linearizable on a per key basis), unlike our work that focuses on providing linearizable transactions across an arbitrary number of keys (potentially entire data set). Our work relies on replicated state machines based on Raft protocol and likewise benefits from its simplicity, theoretical soundness and flexibility (comprehensive semantic).

\end{document}